\documentclass[prb,aps,twocolumn,showpacs,preprintnumbers,amsmath,amssymb,twoside]{revtex4-1}
\usepackage{graphicx}		
\usepackage{pstricks}
\usepackage{dcolumn}		
\usepackage{bm}			
\usepackage{color}
\usepackage{units}		
\usepackage{hyperref}
\hypersetup{colorlinks=true,urlcolor=blue,linkcolor=blue,citecolor=blue}

\newcommand{\LCO}{La$_{{\rm 2}}$CuO$_{{\rm 4}}$}
\newcommand{\LSCO}{La$_{{\rm {2-x}}}$Sr$_{{\rm x}}$CuO$_{{\rm 4}}$}

\begin{document}

\title{Bimagnon studies in cuprates with Resonant Inelastic X-ray Scattering \\
at the O $K$ edge. I - An assessment on La$_2$CuO$_4$ and a comparison with \\
the excitation at Cu $L_3$ and Cu $K$ edges}

\author{V. Bisogni$^{1,2}$}
\altaffiliation[Present address: ]{Leibniz Institute for Solid State and Materials Research IFW Dresden, P.O. Box 270116, D-01171 Dresden, Germany}
\author{L. Simonelli$^1$}
\author{L. J. P. Ament$^3$}
\author{F. Forte$^4$}
\author{M. Moretti Sala$^5$}
\altaffiliation[Present address: ]{European Synchrotron Radiation Facility - Bo\^{i}te Postale 220, F-38043 Grenoble, France}
\author{M. Minola$^5$}
\author{S. Huotari$^{1,6}$}
\author{J. van den Brink$^2$}
\author{G. Ghiringhelli$^5$}
\author{N. B. Brookes$^1$}
\author{L. Braicovich$^5$}

\affiliation{$^1$ European Synchrotron Radiation Facility - Bo\^{i}te Postale 220, F-38043 Grenoble, France}
\affiliation{$^2$ Leibniz Institute for Solid State and Materials Research IFW Dresden, P.O. Box 270116, D-01171 Dresden, Germany}
\affiliation{$^3$ Instituut Lorenz for Theoretical Physics, Universiteit Leiden - P.O. Box 9506, 2300 RA Leiden, The Netherland}
\affiliation{$^4$ CNR-SPIN and Dipartimento di Fisica ``E. R. Caianiello'', Universit\`a di Salerno - I-84084 Fisciano, Salerno, Italy}
\affiliation{$^5$ CNR-SPIN, Dipartimento di Fisica, Politecnico di Milano - Piazza Leonardo da Vinci 32, 20133 Milano, Italy}
\affiliation{$^6$ Department of Physics, P.O. Box 64, University of Helsinki, FI-00014 Helsinki, Finland}

\date{Received: \today}

\begin{abstract}
We assess the capabilities of magnetic Resonant Inelastic X-ray Scattering (RIXS) at the O $K$ edge in undoped cuprates by taking \LCO~ as a benchmark case, based on a series of RIXS measurements that we present here. By combining the experimental results with basic theory we point out the fingerprints of bimagnon in the O $K$ edge RIXS spectra. These are a dominant peak around 450 meV, the almost complete absence of dispersion both with $\pi$ and $\sigma$ polarization and the almost constant intensity vs. the transferred momentum with $\sigma$ polarization. This behavior is quite different from Cu $L_3$ edge RIXS giving a strongly dispersing bimagnon tending to zero at the center of the Brillouin zone. This is clearly shown by RIXS measurements at the Cu $L_3$ edge that we present. The Cu $L_3$ bimagnon spectra and those at Cu $K$ edge - both from the literature and from our data - however, have the same shape. These similarities and differences are understood in terms of different sampling of the bimagnon continuum. This panorama points out the unique possibilities offered by O $K$ RIXS in the study of magnetic excitations in cuprates near the center of the BZ.
\end{abstract}
\pacs{78.70.Ck, 78.70.En, 74.72.Cj, 75.30.Ds}

\maketitle

\section{Introduction}
\label{sec1}

In the debate on high T$_c$ superconductivity the role of the magnetic excitations/fluctuations and the interplay between charge and spin degrees of freedom continue to be among the central issues. \cite{Kivelson2003, Vishik2010} Quite recently this debate has been boosted by the application of Resonant Inelastic X-ray Scattering (RIXS) \cite{Kotani2001, Ament2011} which extended the experimental possibilities offered by well established methods as angle resolved photoemission \cite{Damascelli2003} and neutrons.\cite{Tranquada2007} On the experimental side a breakthrough has been the demonstration that it is possible to study magnetic excitations (namely bimagnon) in cuprates with RIXS. \cite{Hill2008} This new approach cross fertilizes especially with neutron spectroscopy actively used in the study of magnons (see recent work in Refs.\, \onlinecite{Vignolle2007, Lipscombe2009, Headings2010,Stock2010}) and expands the experimental possibilities; in fact generally speaking x-rays allow a wide exploration of $q$-space with easy detection of high energy excitations and require very tiny amount of sample material as opposite to neutrons. On the other hand the major drawbacks are the energy resolution which is much worse than with neutrons and the competition with fluorescence which becomes overwhelming in good metals making the application of RIXS to magnetic excitations very difficult or even impossible. For these reasons RIXS is particularly suitable in (quasi) bidimensional cuprates since the low energy scale is expanded by the high superexchange and the RIXS signal is found up to high dopings covering, to say the least, a great fraction of the dopings corresponding to the superconducting dome in the phase diagram. 

Successively it has been shown experimentally \cite{Braicovich2010} and explained theoretically \cite{Ament2009,Haverkort2010} that RIXS at the Cu $L_3$ edge gives direct access to single magnon excitations. Ref.\,\onlinecite{Braicovich2010} shows moreover that in the underdoped regime magnon-like excitations reminiscent of the excitations in the parent compound are seen up to high energies (typically 350 meV). The opportunities offered by RIXS have been recently exploited with Cu $L_3$ RIXS in the study of the YBa$_2$Cu$_3$O$_{6+x}$ family showing that damped magnons (i.e. paramagnons) are present in all cases studied therein spanning from underdoped to slightly overdoped regime.\cite{Tacon2011} Moreover, Ref.\,\onlinecite{Tacon2011} shows that these paramagnons are compatible with very high T$_c$ superconductivity supported by spin fluctuations. As far as higher order spin excitations are concerned a mixture of single magnon and bimagnon has been observed with Cu $L_3$ RIXS,\cite{Braicovich2009} while it has been demonstrated that RIXS at the Cu $K$ edge gives experimental access to bimagnon \cite{Hill2008,Ellis2010} in agreement with the theoretical work of Refs.\,\onlinecite{Brink2007,Forte2008,Ament2007,Nagao2007}. Thus with a suitable choice of the method it is possible to study either single magnon and high order odd excitations or bimagnon and high order even excitations offering altogether a detailed panorama of spin excitations in bidimensional cuprates. Needless to say this possibility is of paramount importance in the debate on superconductivity.

The topic of the present paper and of the successive one is mainly RIXS at the oxygen $K$ edge respectively in a benchmark parent compound, the \LCO, and in the derived high T$_c$ systems, the \LSCO~ family. In the panorama sketched above it might seem that RIXS at the O $K$ edge has nothing or little to add to the above scenario. Here we demonstrate that this is not the case and that O $K$ RIXS offers unique opportunities. It is in fact possible to map the even order spin excitations dominated by an almost non-dispersing bimagnon with a peak around 450 meV in the low $q$ range (up to about 40\% of the BZ boundary). We show that this region of the parameter space is not accessible to Cu $L_3$ and Cu $K$ RIXS while it is known to be, to say the least, unfavorable to neutron scattering. On top of that traditional Raman spectroscopy is limited to the case of zero momentum transfer. The present work is timely because of the rapid evolution of the field and needed because very little on spin excitations seen with RIXS at the oxygen $K$ edge can be found in the literature where a systematic treatment is lacking: in this connection we mention the bimagnon feature measured with O $K$ RIXS in Sr$_2$CuO$_2$Cl$_2$, reported for the first time in 2002 by Harada $et~al.$ \cite{Harada2002} and more recently by Guarise $et~al.$ \cite{Guarise2010}

Our results on O $K$ RIXS are presented in two twin papers. Here (paper I) we assess the method and by comparing with RIXS at the other edges we obtain a comprehensive overview of bimagnon spectroscopy. To this end we present also original measurements at the Cu $K$ and $L_3$ edges. In the spirit of a mostly phenomenological paper we take advantage of a basic theoretical approach based on Ref.\,\onlinecite{Forte2008}. A byproduct is the way of disentangling mono and bimagnons in Cu $L_3$ spectra. In paper II we exploit the peculiarities of O $K$ RIXS to study a typical superconductor family i.e. the doped \LSCO. Thus the present paper identifies the fingerprints of even order spin excitation seen at oxygen $K$ edge. 

The present paper is organized as follows. We give the experimental details for RIXS at the three edges in Sec.\,\ref{sec2}. The O $K$ data are given in Sec.\,\ref{sec3}, where they are presented (subsection \ref{sec3a}) and discusssed (subsection \ref{sec3b}). Next, in Sec.\,\ref{sec4} data measured at Cu $L_3$ edge and Cu $K$ edge are given with a comparison between O $K$ and Cu $L_3$ (in subsection \ref{sec4a}) and between Cu $K$ and Cu $L_3$ (in subsection \ref{sec4b}). The conclusions are drawn in Sec.\,\ref{sec5}. The paper is supplemented by two appendices: Appendix\,\ref{appA} presents the method to extract the bimagnon excitation at the Cu $L_3$ edge; Appendix\,\ref{appB} deals with multiphonons contributions at the O $K$ edge.

\section{Experimental}
\label{sec2}

\begin{figure}[t!]
\center{
\resizebox{0.85\columnwidth}{!}{%
 \includegraphics[clip,angle=0]{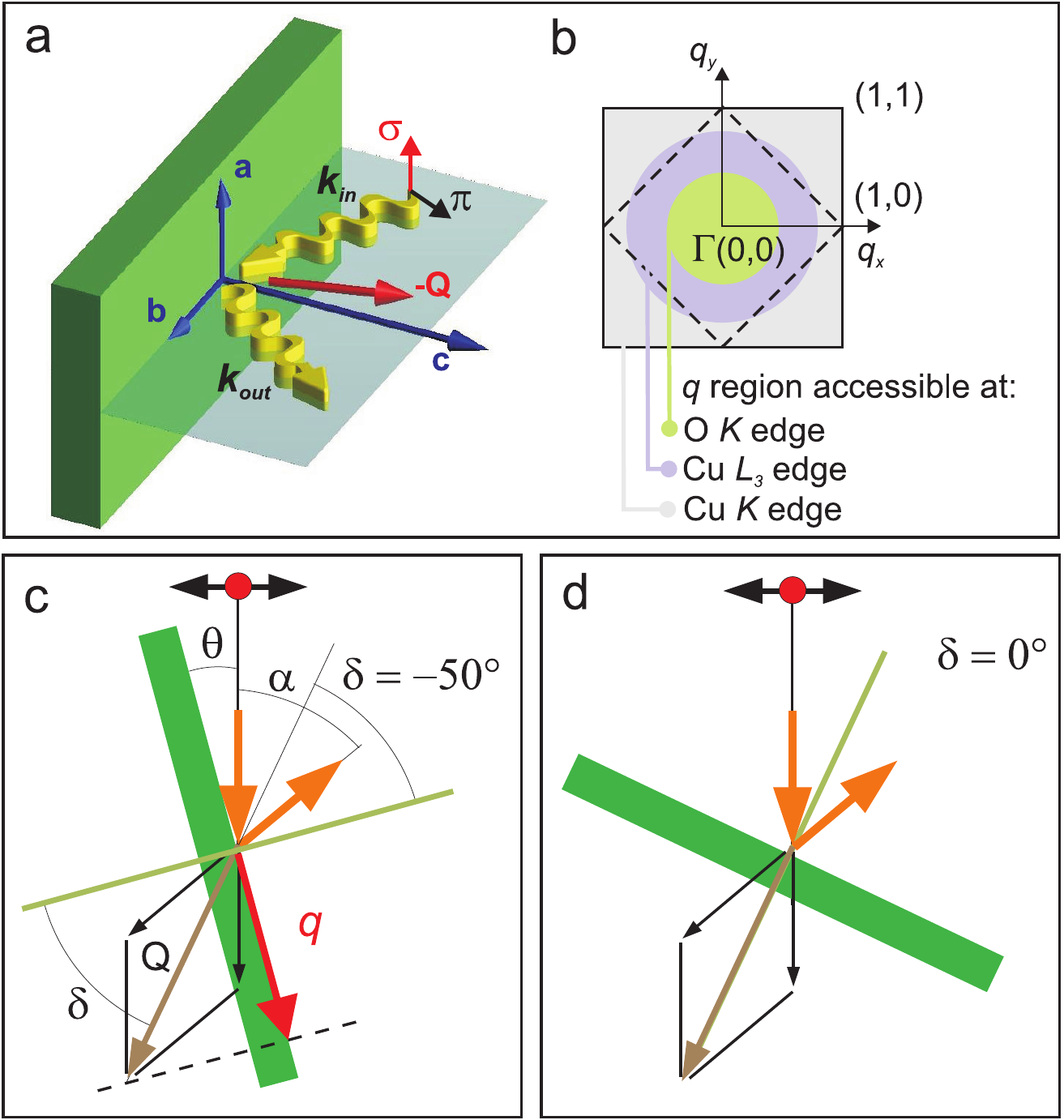}}}
\caption{(Color online). a) The geometry of the RIXS experiment: the sample is in green with its crystallographic axis in blue, while the scattering plane is in light blue. The red and the black arrows indicate the incoming photon polarizations, respectively parallel ($\sigma$) and perpendicular ($\pi$) to the (a,b) plane. b) The cuprate Brillouin zone. The dashed line marks the magnetic BZ boundary. The colored area highlight the $q$ region of the 1$^{\mathrm{st}}$ BZ accessible at the O $K$ edge in the present mounting (green), the Cu $L_3$ edge (lilac) and the Cu $K$ edge (orange). In c) and d) the two different experimental geometries correspond to the maximum momentum transfer condition ($\delta=-50^{\circ}$) and the zero parallel momentum transfer condition ($\delta=0^{\circ}$).}
\label{scat_geom}
\end{figure}

The \LCO~(LCO) sample used in soft x-ray RIXS at the O $K$ and Cu $L_3$ edges was a 100 nm film grown by pulsed laser deposition on (001) SrTiO$_3$, and is the same used in Ref. \onlinecite{Braicovich2010}. The sample measured with hard x-rays at the Cu $K$ edge was a single crystal grown by travelling floating zone method.\cite{Chang2008} In both cases the surface was the basal $(a,b)$ plane with the $c$ axis in the scattering plane (see Fig.\,\ref{scat_geom}).

The O $K$ RIXS measurements (photon energy $\sim$ 530 eV) were done at the beamline ID08 of the ESRF equipped with the AXES spectrometer;\cite{ccd} the measurements were taken in part before and in part after an intervention on the equipment. In the first phase the AXES spectrometer was looking (with 110$^\circ$ scattering angle) to the sample before the entrance slit of the Dragon monochromator of the beamline and the incident monochromatic light was given by the dedicated Polifemo monochromator.\cite{polifemo} In the second phase AXES was sitting at the end of the beamline past the Dragon monochromator with a sagittal refocusing done by a toroidal mirror. In this last case the scattering angle is 130$^\circ$. The combined energy resolution was adjusted to the same value (around 150 meV) in both setups and the new installation resulted in a 3-4 times higher count rate. In the present paper the spectra taken with the second set up are used unless otherwise stated. The average time per spectrum with the Dragon set-up was typically 4 hours for $\sigma$ polarization i.\,e. with the electric vector in the $(a,b)$ plane. 

The Cu $L_3$ RIXS measurements were done at the ADRESS beamline \cite{Strocov} of the Swiss Light Source equipped with the SAXES spectrometer,\cite{saxes} with a scattering angle of 130$^\circ$, and a combined energy resolution of 130 meV at 930 eV. Each spectrum took 30 min. A summary of the geometrical configuration used for experiments in the soft x-rays is given in Fig.\,\ref{scat_geom}(a). The total momentum transfer is $\bf Q=k_{in}-k_{out}$. Being the sample basically bidimensional, the interesting component of the momentum is given by $q$, the projection of {\bf Q} onto the basal $(a,b)$ plane. This is controlled by choosing the scattering angle $\alpha$ and by rotating the sample with $\theta$. Hereafter the angular rotation is indicated by the angle $\delta$ between the sample surface normal and the bisector of the scattering angle. Thus $q \propto sin\delta$ (see Fig.\,\ref{scat_geom}(c-d)).

The measurements at Cu $K$ edge were done at ID16 of the ESRF equipped with the RIXS spectrometer of Refs. \onlinecite{Huotari2006,Huotari2005}, in Rowland circle geometry. The experimental configuration was chosen according to Ref. \onlinecite{Hill2008}: we used only $\pi$ type polarization in order to maximize the projection of the incoming photon wave vector onto the $c$ axis (at this edge the bimagnon is observed only if the incoming polarization has a component along that axis \cite{Hill2008}) and a scattering angle close to 90$^{\circ}$ to reduce the elastic contribution. The spectra were measured near the second Brillouin zone (BZ) indexed (2,2,11) in order to avoid diffraction peaks and to have the reduced momentum transfer $q$ completely defined in the ($a,b$) plane. The combined energy resolution was adjusted to 170 meV for this experiment. The average acquisition time per spectrum was 8 hours. In the following, the Cu $K$ spectra will be presented versus the reduced momentum $q$ reported to the first BZ. All measurements at all edges were taken at 20 K.

\section{The bimagnon at O $K$ edge}
\label{sec3}

\begin{figure}[t!]
\center{
\resizebox{1.05\columnwidth}{!}{%
\includegraphics[clip,angle=0]{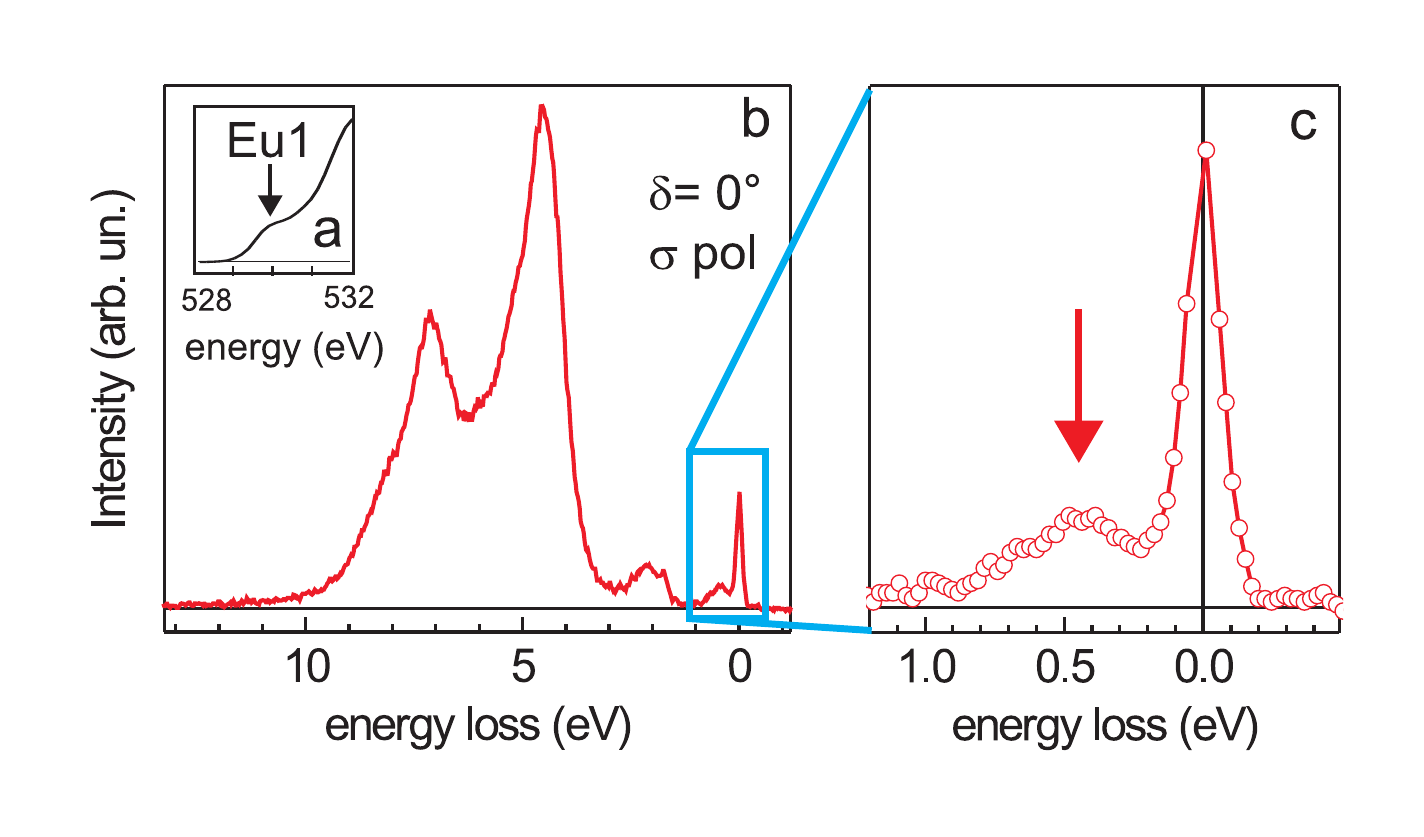}}}
\caption{(Color online). a) O $K$ absorption. b) Emission spectrum of \LCO~ measured with $\sigma$ polarization at $\delta=0^{\circ}$. The excitation energy, $Eu1$, is marked by the arrow in a). c) The low energy expansion of the spectrum shows the bimagnon feature. The red arrow points to its peak position, around 450 meV.}
\label{lco_overview}
\end{figure}

The RIXS study presented here is carried out with excitation in the threshold region of the oxygen $K$ edge absorption spectrum. The absorption at this edge has been studied extensively, \cite{Chen1991,Chen1992,Peets2009} because it sheds light into the valence states hybridized between O $2p$ and Cu $3d$ states. This is the origin of the pre-edge peak shown by the arrow (labeled $Eu1$) in Fig.\,\ref{lco_overview}a and is characteristic of undoped cuprates. This excitation is used in the present RIXS work, which is indeed a resonant study since, as we will show, there are spectral features suppressed or drastically reduced with the excitation energy above the peak. The general shape of the emission spectrum is given in Fig.\,\ref{lco_overview}b. As it is known most of the intensity belongs to the resonant fluorescence\cite{Guo1994,Butorin2000} which covers the range between about 3 eV and 10 eV energy loss when the sample is excited as in Fig.\,\ref{lco_overview}. Moreover the broad peak around 2 eV is the superposition of $dd$-excitations \cite{Moretti2011} on the Cu site and of charge-transfer exciton of the kind of a Zhang-Rice singlet.\cite{Harada2002,Ellis2008} At even smaller energy loss i.e. in the Mid Infrared (MI) range (blue box in Fig.\,\ref{lco_overview}b expanded in Fig.\,\ref{lco_overview}c), the spectrum shows the feature we are interested in. In this region we see a dominant peak around 450 meV whose dependence on excitation energy, incident polarization and transferred momentum is presented hereafter. We will show that all these results are consistent with the assignment of this peak to a bimagnon excitation together with higher order spin excitations of the same symmetry i.e. with an even number of spin flips. When no misunderstanding is possible we will simplify the language by calling conventionally bimagnon the experimental result including also the higher orders.

\subsection{Experimental results in the Mid Infrared}
\label{sec3a}

\begin{figure*}[t!]
\center{
\resizebox{1.86\columnwidth}{!}{%
\includegraphics[clip,angle=0]{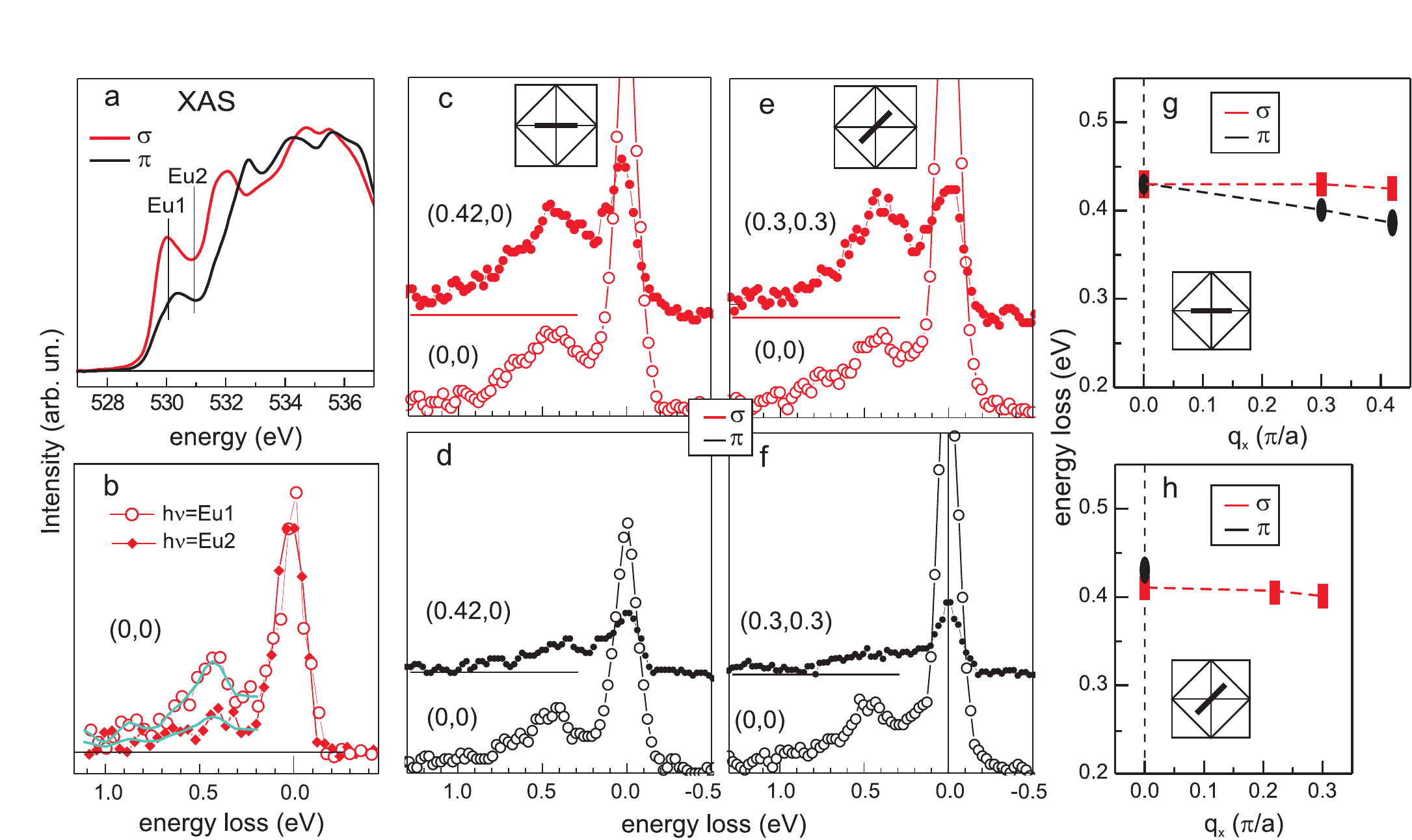}}}
\caption{(Color online). a) XAS spectrum of \LCO~ for $\sigma$ (red) and $\pi$ (black) polarizations ($\delta=-45^{\circ}$) measured in fluorescence yield mode. b) On resonance (red open dots) and out of resonance (red filled diamond) RIXS spectra measured respectively at $Eu1$ and $Eu2$ excitation energies, marked in a). The blue line is a smoothing of the inelastic part, as a guide to the eye. c-f) O $K$ RIXS spectra measured at $Eu1$ as a function of momentum and polarization (scattering angle 130$^\circ$). c) and e) RIXS spectra along the (1,0) direction respectively for $\sigma$ (red) and $\pi$ polarization (black); d) and f) refer to the (1,1) direction. Spectra for $q \neq 0$ (filled symbols) are vertically offset with respect to the spectra for $q$=0 (open symbols). All the spectra have been normalized to the acquisition time. g-h) Summary of the bimagnon dispersion respectively along the (1,0) and along the (1,1) direction. The data are expressed as a function of q$_{\rm x}$, the component of the in-plane momentum $q$ along the (1,0) direction. Data points for $q$=(0.3,0) and (0.21,0.21) refer to the spectra measured at 110$^{\circ}$ scattering angle.} 
\label{lco_qdep}
\end{figure*}

The main experimental results are summarized in Fig.\,\ref{lco_qdep}. The panel \ref{lco_qdep}a shows the polarization dependence of the absorption; note that the spectra are consistent with the literature.\cite{Chen1991,Chen1992,Peets2009} 
The excitation energy dependence is given in panel \ref{lco_qdep}b showing the resonant behavior of the MI feature at the peak excitation $Eu1$ with a drastic decrease at excitation energy $Eu2$, above the peak (data taken at 110$^\circ$ scattering angle). The excitation energies are given by the vertical bars in panel \ref{lco_qdep}a. 

The polarization and the momentum dependence of RIXS are shown in panels \ref{lco_qdep}c-f where the black spectra give the $\pi$ incident polarization and the red spectra the $\sigma$ one (data taken at 130$^{\circ}$ scattering angle). These are raw data normalized to the acquisition time. These results show two characteristic aspects. First the dispersion with the transferred moment $q$ is, if any, very small both along the direction from $\Gamma$ to (1,0) and from $\Gamma$ to (1,1); this is summarized in panels (g) and (h) including also the results at 110$^{\circ}$ scattering angle (q=(0.3,0) along (1,0)). A weak but visible dispersion is seen only with $\pi$ polarization along the (1,0) direction. Second the spectral weight of the MI feature is basically determined by the absorption, i.e. the excitation channel with $\pi$ polarization has a typical loss of intensity at grazing incidence (high $q$). In this case the electric vector has a small component in the (a,b) plane so that the probability of exciting the system decreases, being the ground state of $x^2-y^2$ type with the holes in the (a,b) plane.

\subsection{Discussion}
\label{sec3b}

The assignment of the MI excitations was the object of long standing discussions on the relative weights of polaronic effects and of magnetic excitations;\cite{Kastner1998,Lorenzana1995} now the situation is settled at least in RIXS at Cu $L_3$ and $K$ edges with a compelling evidence of the assignment to magnetic excitations. As a matter of fact at Cu $K$ edge the lowest magnetic excitation is the bimagnon since there is no spin-orbit so that only an even number of spin flips besides no spin flip is allowed; in this connection the small spin orbit in the valence state of $3d$ systems has negligible effect. On the other hand at Cu $L_3$ edge the core hole spin-orbit is large and the spin is not a constant of the motion; thus the single magnon is clearly seen with its characteristic dispersion in agreement with the neutron results. Obviously this framework and the analogy with Cu $K$ edge indirect excitation of bimagnon is a strong support to the assignment of the MI feature in O $K$ RIXS to a bimagnon. We stress that in both cases there is no spin-orbit. The observation in Sr$_{ 2}$CuO$_2$Cl$_2$ at O $K$ edge and assignment to bimagnon presented at a single $q$ value by Guarise $et$ $al.$ \cite{Guarise2010} is essentially based on these qualitative arguments. Well before this work a faint feature has been seen and correctly assigned in Sr$_{ 2}$CuO$_2$Cl$_2$ by Harada $et$ $al.$ \cite{Harada2002} but the experimental basis was at that time very limited due to technical constraints.

\begin{figure}[t!]
\center{
\resizebox{0.95\columnwidth}{!}{%
\includegraphics[clip,angle=0]{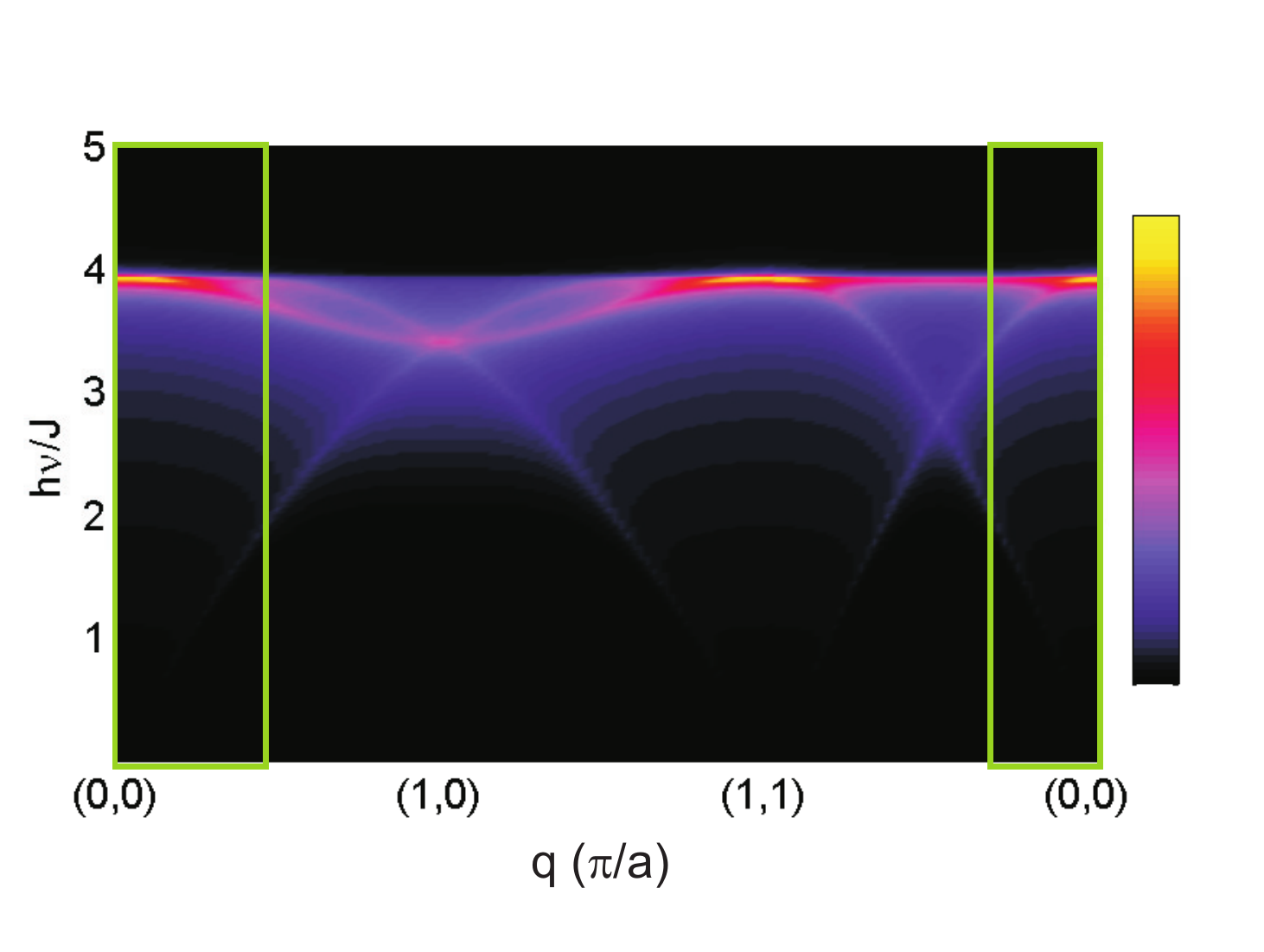}}}
\caption{(Color online). The calculated two-magnon density of states for nearest-neighbor interaction and without magnon-magnon interaction, as reported in Ref. \onlinecite{Forte2008}. The green rectangles highlight the BZ portions investigated at the O $K$ edge. The energy scale on the left is in units of $J$, while the one on the right is in eV ($J$=135 meV \cite{Coldea2001}).}
\label{dos}
\end{figure} 

A first strong support to this assignment is the comparison of the RIXS results with the density of states (DOS) of the bimagnon; to this end the non-interacting results of Fig.\,\ref{dos} (reproduced from Ref.\,\onlinecite{Forte2008}) are sufficient. The O $K$ RIXS process covers a limited region of $q$ around the $\Gamma$ point. In the figure the two green boxes give the regions accessible at the O $K$ edge due to the limited available momentum. In this range the DOS has a prominent intensity at high energy and a weak intensity at very low energy with a continuum in between. Unless dramatic effect of the matrix elements comes into play, the sampling of the bimagnon would bring to a dominant feature at high energy. 
This would be somewhat lower than 4$J$, where $J$ is the superexchange. Being this calculation without interaction, it is expected that the energy scale is overestimated with respect to the experiment, as it will be discussed in the following.
With this sampling the absence of dispersion or a small dispersion are a quite natural consequence.

\begin{figure}
\center{
\resizebox{0.96\columnwidth}{!}{%
\includegraphics[clip,angle=0]{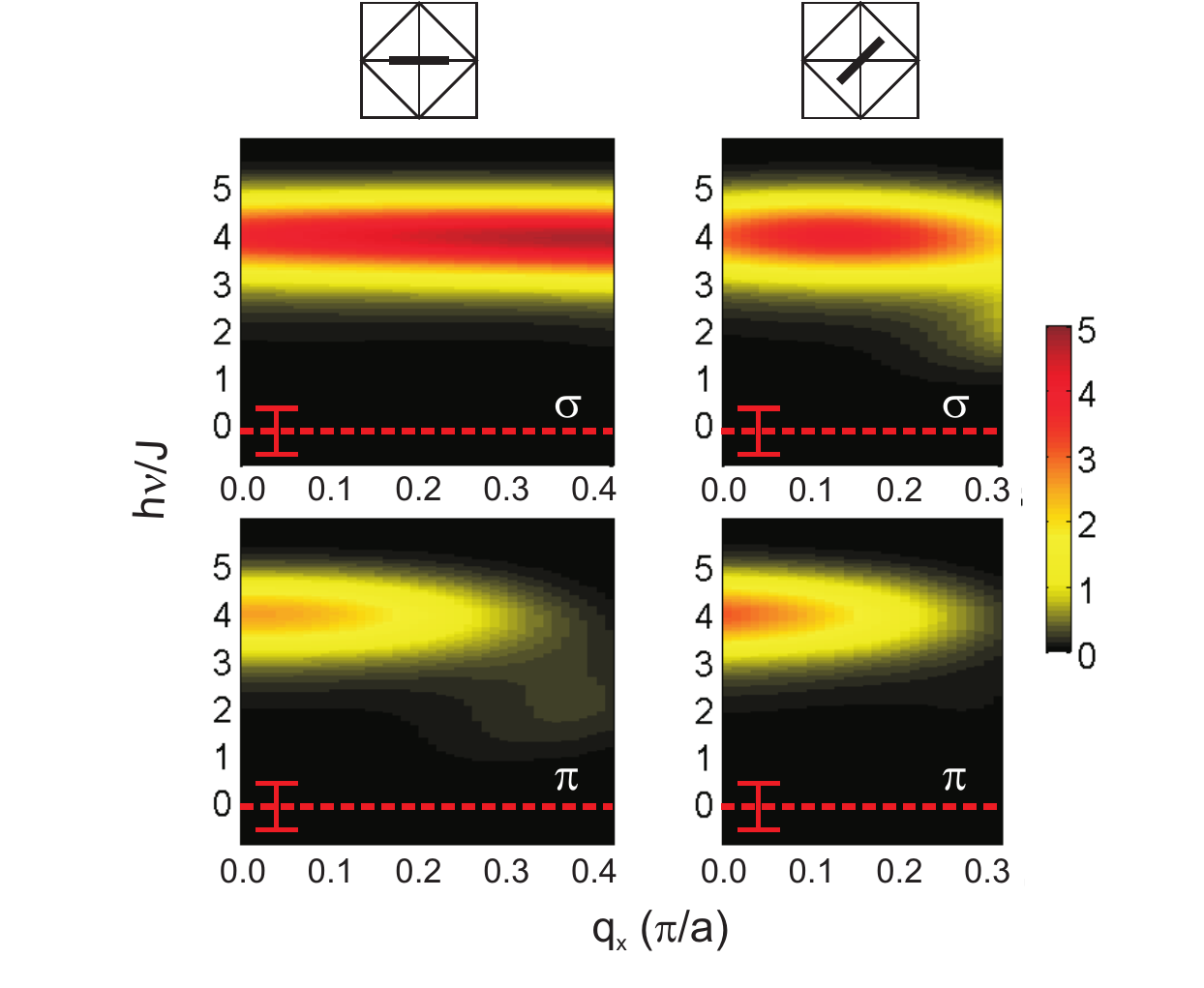}}}
\caption{(Color online). RIXS theoretical cross sections calculated as in Ref.\,\onlinecite{TAment} with 130$^\circ$ scattering angle. The experimental broadening and self-absorption correction are included. The columns refer to the inspected BZ directions, (1,0) and (1,1), while the rows refer to the polarizations $\sigma$ and $\pi$. The red vertical bars indicate the instrumental resolution, $\sim~J$. As in Fig.\,\ref{lco_qdep}, q$_{\rm x}$ is the component of the in-plane momentum $q$ along the (1,0) direction.}
\label{cross_section}
\end{figure}

Going into more details the assignment to the bimagnon is supported also by a basic theoretical model already used in the indirect bimagnon excitation at Cu $K$ edge.\cite{Forte2008} This is based on the linear term in a series of powers of $J/\Gamma$ representing the scattering operator ($\Gamma$ is the core level lifetime broadening). With respect to Ref.\,\onlinecite{TAment} one introduces the oxygen ions and the hopping from O $2p$ to Cu $3d$. Also in this case, in order to understand the type of sampling, the inclusion of magnon-magnon interaction is not needed. The theory confirms that one is sampling the upper intense ridge of the DOS. This is shown by the comprehensive view of the theoretical cross sections corrected for self-absorption and instrumental resolution given in Fig.\,\ref{cross_section} by color maps in nice qualitative agreement with the experimental results. In fact basically no dispersion is seen with $\sigma$ polarization while a tiny dispersion might be compatible with the $\pi$ excitation along (1,0). Also the trend of the intensities is well reproduced by the theory. As a matter of fact along the (1,0) direction with $\sigma$ polarization also the small increase of the intensity seen in the experiment is obtained. This is due to the reduction of self-absorption at grazing incidence and is an effect of about 20\%; note that this is the only case in which the self-absorption correction really matters. 

Having assessed the O $K$ edge RIXS it is interesting to compare with the other edges at Cu. This is done in the next section.

\section{Comparison with the other edges}
\label{sec4}

In order to obtain a general view of bimagnon spectroscopy, another piece of experimental information is missing at present. This is the bimagnon information one obtains from Cu $L_3$ edge RIXS. Indeed the work at Cu $L_3$ edge is emerging as a powerful tool in the study of single magnon. As a consequence, up to now minor attention has been given to the information on bimagnon basically left dormant in the raw data. Here we present results on the bimagnon at the Cu $L_3$ edge in a wider context i.e. the comparison between the spectra of bimagnon at three edges i.e. O $K$, Cu $L_3$ and Cu $K$. This is an interesting problem since at each $q$ value, the bimagnon density of states is represented by a continuum so that different sampling are typical of RIXS at each edge and there is no a priori reason why they should be equal. To make this program viable we need a way to extract the bimagnon information from the $L_3$ data where it is superimposed to single magnon. More precisely we need a method to separate the contributions from odd and even number of spin flips. Also this problem is addressed in the present paper and for better readability this methodological contribution is presented separately in Appendix \ref{appA}.

\subsection{Comparison between O $K$ and Cu $L_3$ RIXS}
\label{sec4a}

\begin{figure}[t!]
\center{
\resizebox{\columnwidth}{!}{%
\includegraphics[clip,angle=0]{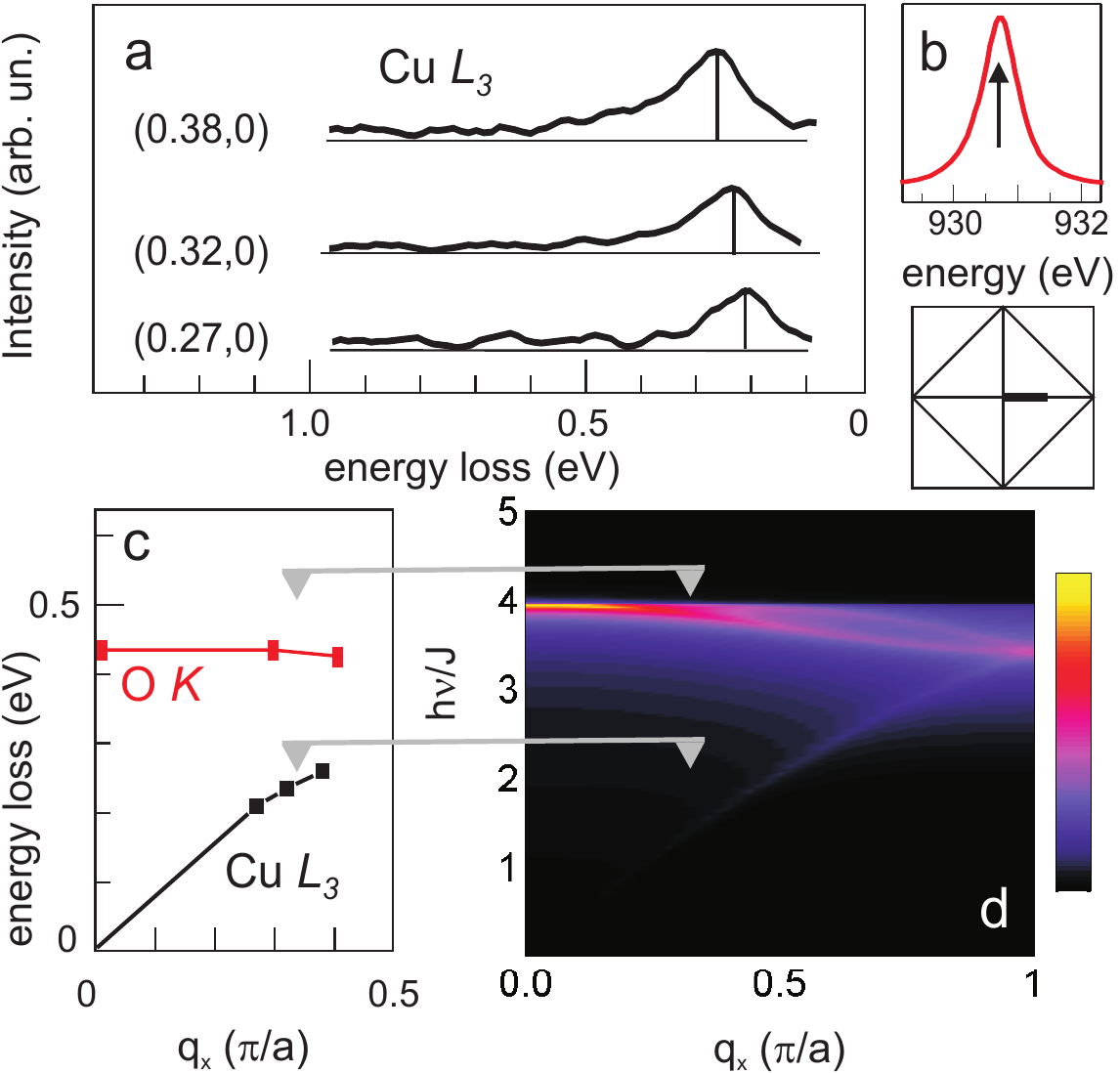}}}
\caption{(Color online). a) The bimagnon extracted from the Cu $L_3$ RIXS spectra along the (1,0) direction with $\sigma$ polarization. b) Cu $L_3$ XAS of LCO. c) The dispersion as seen with O $K$ edge (red squares) and Cu $L_3$ (black squares) for the (1,0) direction. The gray arrows point out the correspondence to the theoretical two-magnon DOS along (1,0), in d).\cite{Forte2008}}
\label{cu_dos}
\end{figure}

This comparison is done with $q$ along the direction from $\Gamma$ to (1,0) and the main results are summarized in Fig.\,\ref{cu_dos}. In panel \ref{cu_dos}a we give the bimagnon spectra extracted from the $L_3$ data with the method given in App.\,\ref{appA}; the excitation is at the $L_3$ peak in the absorption as specified in panel \ref{cu_dos}b. Note that the bimagnon spectra extracted as in Appendix \ref{appA} have the same spectral function independently of the polarization which determines simply a scaling factor. The results of panel \ref{cu_dos}c show a clear positive dispersion at $L_3$ as opposite to the small negative dispersion at higher energies seen in O $K$ data and already discussed. This striking difference is understood once again in terms of different sampling of the bimagnon continuum as shown in the comparison with the density of states (panel \ref{cu_dos}d). The correspondence of the Cu $L_3$ results with the lower ridge of bimagnon continuum is very clear (as usual the energies in the density of states are in excess since the magnon-magnon interaction is neglected). This means that a particular subset of states is emphasized at each edge by matrix elements and not that some states are rigorously excluded due to a selection rule. In fact the calculations at the oxygen edge show a very tiny but finite intensity in correspondence to the lower ridge of the density of states. The above analysis cannot be extended to much lower $q$ values than in Fig.\,\ref{cu_dos} due to the difficulty of disentangling not only bimagnon from single magnon but also from phonons and multiphonons. In this connection we have the evidence that at the oxygen edge there is also a multiphonon contribution. This is not the topic of the present paper and for completeness is briefly summarized in Appendix \ref{appB}. 

\subsection{Comparison between Cu $K$ and Cu $L_3$ RIXS}
\label{sec4b}

\begin{figure}[t!]
\center{
\resizebox{0.95\columnwidth}{!}{%
\includegraphics[clip,angle=0]{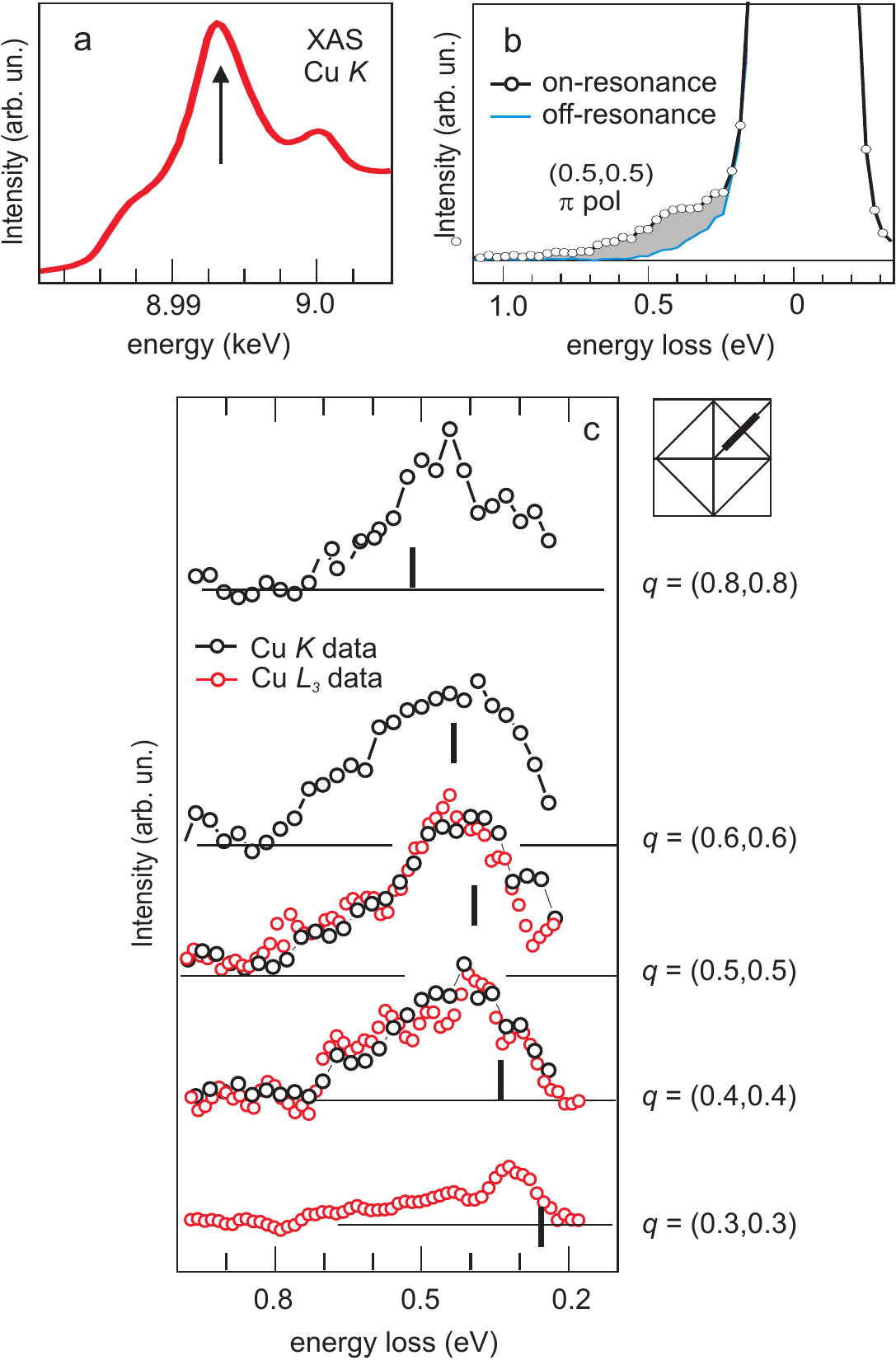}}}
\caption{(Color online). a) Cu $K$ XAS of LCO. b) Typical RIXS spectrum with Cu $K$ excitation at the energy marked in a): the gray area is the bimagnon while the blue line indicates the out of resonance elastic background measured 40 eV below the main XAS peak. c) Cu $L_3$ and Cu $K$ bimagnon along (1,1) direction, respectively in red and black dots. The vertical offset between two consecutive spectra is proportional to the difference of the corresponding momenta (marked on the right). The black ticks mark the theoretical bimagnon peak positions obtained from theory as in Refs. \,\onlinecite{Forte2008,TAment} with $J$=135 meV.\cite{Coldea2001}}
\label{cu_bimagnon}
\end{figure}

While the spectroscopy of magnetic excitations at Cu $K$ has been investigated rather extensively in the last few years,\cite{Hill2008,Ellis2010} the comparison between $L_3$ and $K$ edges in terms of bimagnon spectral function remains an open issue. Indeed we have just shown that Cu $L_3$ explores the lower ridge of the bimagnon continuum; this type of sampling is done also at the Cu $K$ edge accordingly to Ref.\,\onlinecite{Forte2008} so that a similarity between Cu $K$ and Cu $L_3$ is expected. But similarity does not mean necessarily identity. In effect within the approximation used in Ref.\,\onlinecite{Forte2008} the two cases should give the same spectral function apart a scaling factor, but this is only an approximate scheme and in the real world there might be differences. Therefore an experimental comparison between RIXS at the two edges is significant and is given here. In this connection it is useful to have $L_3$ and $K$ spectra exactly at the same $q$ values and to this purpose we measured also Cu $K$ spectra. These agree with the literature in the cases already investigated and add new information. 
As it is well known, in the Cu $K$ edge spectra the bimagnon is a small feature on top of the tail coming from the elastic peak as in Fig.\,\ref{cu_bimagnon}b which gives a spectrum measured at the Cu $K$ edge absorption peak (Fig.\,\ref{cu_bimagnon}a). The Cu $K$ RIXS spectra are given by the black open dots in Fig.\,\ref{cu_bimagnon}c; these are raw data after background subtraction and not rescaled so that the relative heights of the spectra are significant. To these results the extracted $L_3$ spectra are superimposed (open red dots); also in this case the relative heights are preserved while the heights of the two sets ($K$ and $L_3$) are adjusted to superimpose the spectra at (0.6,0.6) and at (0.5,0.5) (units so that the corner of the BZ is at (1,1)). The main results is that the spectral shape is the same at the two edges within the experimental errors dictated by statistics, which is on the average 20\%.

It has been already demonstrated by Ellis $et$ $al.$ \cite{Ellis2010} that the accurate fitting of the bimagnon energy requires the inclusion of the magnon-magnon interaction so that there is no point in doing this exercise in the present case. Instead it is interesting to test at Cu $K$ edge the basic approach we used in the O $K$ case, with the obvious needed changes.\cite{Forte2008,TAment} This approach at lowest order in $J/\Gamma$ with only nearest neighbor interaction (i.e. without second and third neighbors and without ring exchange) gives the maximum of the bimagnon shown by the black ticks in Fig.\,\ref{cu_bimagnon}. These are obtained with $J$= 135 meV from Ref.\,\onlinecite{Forte2008}. As expected the calculation overestimates the energy at higher $q$. Moreover there is some under-estimation at low $q$ but the general picture is remarkably good. This is a further, although indirect, support to the use of a conceptually equivalent approach at O $K$ edge. In this analysis we refer to the peak position which is a safe indicator only marginally affected by the higher order contributions, clearly present in the measurements. In fact the end point of the measured spectra is around 1 eV while the end point of the bimagnon with interaction and including the experimental resolution cannot exceed about 0.6 eV.

\section{Conclusions}
\label{sec5}

We have presented experimental results on Mid Infrared excitations in RIXS at the O $K$ edge in the benchmark case of undoped LCO and we have shown that the spectrum is given by the even order magnetic excitations and in particular is dominated by bimagnon. On this basis we have pointed out the fingerprints of bimagnon excitation which are available to be used to identify bimagnon also in more complex systems by using O $K$ RIXS. The most typical aspect is the almost complete absence of dispersion and another aspect is the roughly constant intensity with polarization vs. $q$ (going toward grazing incidence) when incoming $\sigma$ polarized light is used. The behavior of the intensity is better seen by including corrections for self-absorption which however do not change the general picture. These typical results are understood by comparing directly with the non-interacting bimagnon density of states. This shows that O $K$ RIXS is sampling the upper ridge of the density of states near the $\Gamma$ point; this argument is also supported by a basic theoretical treatment of the type of Ref.\,\onlinecite{TAment}.

Moreover a more general picture of bimagnon spectroscopy is obtained by comparing the O $K$ RIXS with Cu edges results i.e. $L_3$ and $K$ RIXS. Both Cu edges explore the lower ridge of the bimagnon density of states and show clear dispersion. In the $L_3$ case we present original data obtained with a separation procedure of odd and even order excitations. This allows also a comparison between $L_3$ and $K$ edge data taken to this purpose; in both cases the spectral functions have the same shape within the noise, a point never assessed before.

\begin{table*}
\begin{center}
\begin{tabular}{ p{5 cm} p{3.3 cm} p{3.3 cm} p{3.3 cm} }
\hline
\hline\noalign{\smallskip}
 & {\bf O $K$} & {\bf Cu $L_3$} & {\bf Cu $K$} \\
\hline\noalign{\smallskip}
{\bf transition} & $1s\rightarrow 2p$ & $2p_{3/2}\rightarrow 3d$ & $1s\rightarrow 4p$ \\
\hline\noalign{\smallskip}
{\bf typical energy} & 530 eV (soft x-ray) & 930 eV (soft x-ray) & 8990 eV (hard x-ray) \\
\hline\noalign{\smallskip}
{\bf core-hole spin-orbit} & no & yes & no \\
\hline\noalign{\smallskip}
{\bf spin excitation} & $\Delta$S=0 & $\Delta$S=0,$\pm 1$ & $\Delta$S=0\\
\hline\noalign{\smallskip}
{\bf single magnon} & absent & dominant & absent\\
\hline\noalign{\smallskip}
{\bf physical information} & upper ridge bimagnon & lower ridge bimagnon & lower ridge bimagnon\\
{\bf on the bimagnon} & (almost flat) & (dispersing) & (dispersing)\\
\hline\noalign{\smallskip}
{\bf momentum extension} & $\sim$ 45\% of the 1$^{\mathrm{st}}$ BZ & $\sim$ 80\% of the 1$^{\mathrm{st}}$ BZ & several BZs \\
\hline\noalign{\smallskip}
{\bf RIXS contribution with} & & & \\
{\bf respect to total emission} & {small} & {large} & {small} \\
\hline\noalign{\smallskip}
{\bf data handling needed} & & delicate separation & delicate background \\
{\bf to obtain bimagnon} & straightforward & from single magnon & subtraction \\
\hline
\hline
\end{tabular}
\caption{Summary of the main aspects of RIXS at the three edges O $K$, Cu $L_3$ and Cu $K$.}
\label{t1}
\end{center}
\end{table*} 

The difference in the sampling of the DOS given by oxygen and copper sites is not due to a selection                                                                             rule but to a drastic difference in the contributions to the spectra from the two ridges. The high energy one dominates the DOS and thus the spectrum as in O $K$ RIXS, unless something special happens to reduce its weight. This is the case of copper excitation where the matrix element effect, as explained in Ref.\,\onlinecite{Forte2008} for Cu-K edge, almost suppresses the upper ridge and in the meantime the total spectral weight tends to zero with decreasing $q$. This last effect comes from the commutation of the first order scattering operator  with the ground state Hamiltonian at $q$=0. For symmetry reasons this happens also at Cu $L_3$, but not at O $K$ edge.\cite{TAment}  

This general view shows that each edge has specific features summarized for convenience in Table \ref{t1}. In particular we stress that O $K$ RIXS although with a limited momentum gives unique information. In fact a sampling of even spin flip excitations near $\Gamma$ point with energies around 450 meV is not possible with $K$ and with $L_3$ RIXS. Thus O $K$ RIXS has a specific role in the study of the low energy scale excitations in Cuprates. 

\acknowledgments 
The work took enormous advantage from numerous stimulating and productive discussions with many colleagues and in particular with B. B\"uchner, J. Geck , J. Mesot and G. Monaco. The Cu $L_3$ RIXS was performed at the ADRESS beam line of the SLS (PSI) using the SAXES spectrometer developed jointly by Politecnico di Milano, SLS, and EPFL. For the measurements taken at the SLS we acknowledge the assistance by T. Schmitt. The authors are grateful to F. Miletto Granozio and M. Radovic for providing the LCO thin film sample and to J. Mesot for the LCO single crystal. V.\,B. acknowledges the financial support from Deutscher Akademischer Austausch Dienst. F.\,F. acknowledges financial support from the FP7/2007-2013 under the grant agreement No.\,
264098-MAMA. S.\,H. was supported by Helsinki University research funds (project number 490076) and the Academy of Finland (project numbers 1256211 and 1254065).

\appendix
\section{BIMAGNON EXTRACTION AT Cu $L_3$}
\label{appA}
The aim of this Appendix is to present the concept on the bimagnon extraction from the spectra at Cu $L_3$. The basis are the transformation properties of the spectral components with the angles and/or the incident polarizations. This makes it possible to separate the odd spin symmetry (single magnon and higher order odd excitations - referred to as the $S$ component) from the even spin symmetry (bimagnon and even higher order - referred to as the $B$ component).

Accordingly to Ref. \onlinecite{Ament2009} the magnetic intensity can be expressed as the product $I = F \times G$ where $F$ is an atomic form factor containing the polarization effect and depending on the excitation one is looking for, while $G$ depends only on the absolute value of the momentum $|q|$. This factorization is essential in the procedure presented here and is possible because the RIXS process has only a single intermediate state; this is an important difference with respect to the more general case presented in Ref. \onlinecite{Haverkort2010}.

We consider the spectra taken at the same absolute value of the transferred momentum $q$ and with the two incident linear polarizations $\sigma$ and $\pi$, i.e. the four cases ($+q,\sigma$), ($+q,\pi$), ($-q,\sigma$), ($-q,\pi$). Moreover we assume they are already corrected for self-absorption. The function $G$ is the same for the quadruplet of spectra so that only the form factor $F$ determines the transformation coefficients of the $S$ and $B$ components within any pair of spectra $I_1$ and $I_2$ arbitrarily extracted from the quadruplet. One can write \\

\begin{center}
$I_1 =  S_1 + B_1$, 
\end{center}
 
\begin{center}
$I_2 = S_2 + B_2 = \gamma S_1 + \beta B_1$
\end{center}
 
so that $I_1$ and $I_2$ can be obtained since the coefficients $\gamma$ and $\beta$ are known from the form factors given in Refs.\,\onlinecite{Braicovich2010a,Moretti2011}. In fact $\gamma = F_{S2}/F_{S1}$. The $B$ symmetry is shared by a variety of channels including the elastic, which has the form factor $F_E$ defined in Ref. \onlinecite{Ament2009}. The factor $F_B$ is proportional to $F_E$ so that $\beta = F_{E2}/F_{E1} = F_{B2}/F_{B1}$.

The decomposition is safe provided that the couple $I_1$ and $I_2$ is properly chosen among the quadruplet. A very convenient choice is to take $I_1$ and $I_2$ at the same angle (i.e. the same $q$) and different polarization (1 = $\pi$ and 2 = $\sigma$). In this case the self-absorption is almost the same and this reduces the uncertainties. 

\section{PHONONS AND OVERTONES}
\label{appB}
Phonon spectroscopy with RIXS is not the topic of the present paper and the instrumentation was not optimized to this end. In spite of that a byproduct is an interesting hint on phonons presented in this Appendix. 

\begin{figure}[t!]
\center{
\resizebox{0.92\columnwidth}{!}{%
\includegraphics[clip,angle=0]{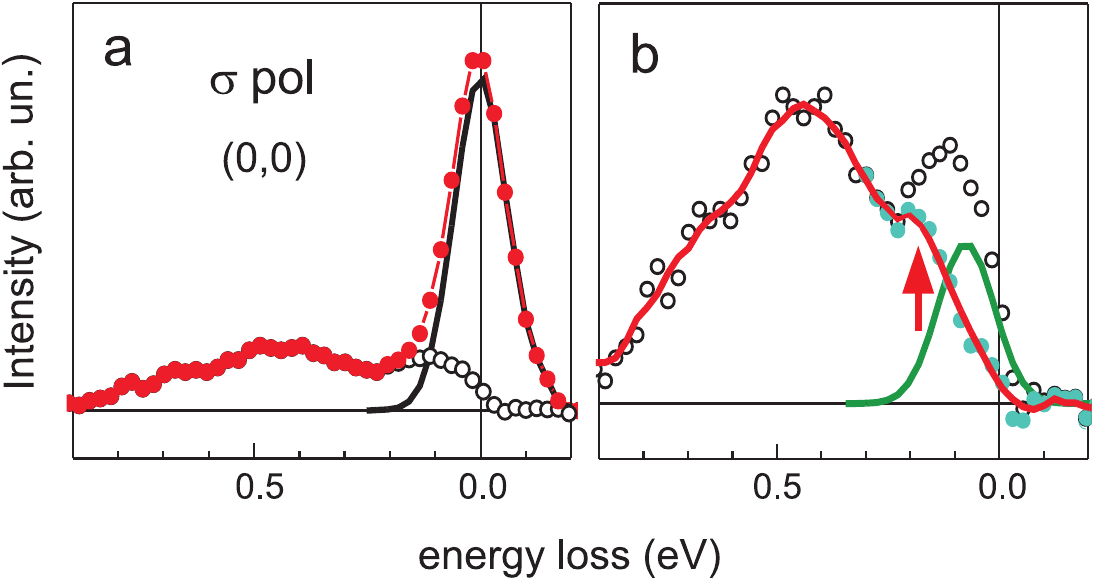}}}
\caption{(Color online). Decomposition of the LCO spectrum (red filled dot line): a) inelastic part (black open dot line) after subtraction of the elastic contribution (black solid line, FWHM=150 meV); b) further decomposition of the low energy inelastic spectrum by subtracting the contribution from the phonon breathing mode at 85 meV (green line). The remaining inelastic spectrum at higher energy losses (blue filled dot) shows the presence of further multiphonon modes (arrow). The red solid line is a smoothing curve as a guide to the eye. }
\label{phonon}
\end{figure}

In the paper we have shown that the bimagnon contribution in the O $K$ RIXS spectra is basically around 3$J$ so that the instrumental resolving power used here is more than enough to study bimagnons. The situation with phonons is quite different: the highest energy phonon (around 85 meV),\cite{Tajima1990} which is also the dominant phonon contribution, is just above the half width (150/2 = 75 meV) of the experimental elastic peak, measured accurately with scattering from a carbon tape. With this resolution the phonon contribution is expected to be seen basically as an asymmetry of the peak around zero energy loss and dominated by the elastic contribution. This is indeed the case as shown in Fig.\,\ref{phonon}(a) based on the data at $q$=0. Once a symmetric elastic peak representing the experimental resolution is subtracted, the difference spectrum (open dots) shows an extra feature around 120 meV. We use a Gaussian elastic peak because this shape is given within the error bars by the carbon tape measurements.

The peak of the difference spectrum belongs to the region comprising phonons and their overtones which are not separated with the present resolution. In effect the high energy phonon has little dispersion so that the biphonon is around 170 meV. A rough decomposition of the open dot spectrum (see Fig.\,\ref{phonon}(b)) is compatible with this argument since the subtraction of a peak at 85 meV gives a reasonable biphonon signature pointed out by the red arrow. This exercise is resolution limited and cannot be quantitative, but it has an important implication besides the obvious fact that better resolution is needed; it shows that with about 2 or 3 times better resolving power the biphonon spectral weight is sufficient to make it observable at the oxygen edge in bidimensional undoped cuprates having a large superexchange. In fact in these systems the magnetic excitations are nearby the biphonons and it is not obvious a priori that the biphonon spectral weight is sufficient. In effect up to now the biphonon has been seen with RIXS at the oxygen edge only in systems having very low $J$,\cite{Bisogni_phonon} where the situation is much more favorable.

As matter of fact the needed resolution is already available, \cite{saxes,Strocov} so that the measurement of the biphonon/phonon ratio in bidimensional cuprates is a real possibility. This is extremely important in the light of the recent theoretical work showing that it is possible to extract directly the e-ph interaction matrix element from this ratio obtained from the RIXS spectra.\cite{Yavas2010}

\end{document}